\documentclass[aps,prd,twocolumn, showpacs, showkeys]{revtex4-1}

\usepackage{graphicx}

\begin{document}


\title{Born$-$Infeld-like modified gravity}

\author{S. I. Kruglov}

\email{krouglov@utsc.utoronto.ca}


\affiliation{Department of Chemical and Physical Sciences, University of Toronto,\\
3359 Mississauga Rd. North, Mississauga, Ontario, Canada L5L 1C6}

\date{\today}

\begin{abstract}
A modified theory of gravity with the function $F(R) = (1-\sqrt{1-2\lambda R})/\lambda$ is
suggested and analyzed. At small value of the parameter $\lambda$ introduced the action is converted into Einstein$-$Hilbert action. The theory is consistent with local tests which gives a bound on the value
of the parameter $\lambda\leq 2\times 10^{-6}$ cm$^2$. I have considered the Jordan frame as well as the Einstein frame in which the potential of the scalar field was obtained. The static Schwarzschild$-$de Sitter solutions of the model are obtained and analyzed. It was demonstrated that the de Sitter space is unstable but a solution with zero Ricci scalar is stable. It is shown that there is no matter instability in the proposed model.
\end{abstract}

\pacs{04.50.Kd, 98.80.Es}

\maketitle

\section{Introduction}

The general theory of relativity (GR) based on the Einstein$-$Hilbert (EH) action can not explain the acceleration of the early and late universe. Therefore, GR does not describe precisely gravity and it is reasonable to modify it in such a way that the theory admits the inflation and imitates the dark energy. In addition, the Mach principle is not included in GR. Thus, the economical strategy is not to introduce extra fields to resolve the cosmological problems but to modify the EH action by introducing proper $F(R)$-gravity. Such theory can describe the early-time inflation and late-time acceleration.
It should be noted that the modified gravity ($F(R)$-gravity) is the phenomenological model describing local tests and observational data. GR was tested in the Solar System in weak-field and slow motion approximation but at the scale of galaxies and clusters and strong gravity
has been tested with poor accuracy. The fundamental theory of relativity has to be
the renormalizable quantum gravity which may describe quantum of gravitational waves \cite{Birrel}.
It should be also mentioned that corrections introduced by renormalization, due to one-loop divergences, contain a scalar curvature squared ($R^2$) as well as Ricci tensor squared ($R_{\mu\nu}R^{\mu\nu}$)
which includes ghosts. Therefore, regardless of the form of the function $F(R)$, $F(R)$-gravity is not
renormalizable.
Nevertheless, $F(R)$-gravity can explain the transition from deceleration to acceleration
(see, for example, \cite{Odintsov}, \cite{Faraoni}).
Note that $F(R)$-gravity contains a single extra degree of freedom (which can be a ghost or a
healthy scalar depending on the chosen function $F(R)$) and can be reformulated in a scalar-tensor form.
We use here the Born$-$Infeld (BI) procedure \cite{Born} to introduce a new dimensional parameter $\lambda$ replacing the Lagrangian density ${\cal L}$ by $(1-\sqrt{1-2\lambda {\cal L}})/\lambda$. It is known that such deformation in electrodynamics allows us to remove divergences connected with point-like charges and to obtain finite self-energy. In this paper the curvature $R$ in the EH action is substituted by $(1-\sqrt{1-2\lambda R})/\lambda$ with new scale $\lambda$.

The paper is organized as follows. In Sec.2, a model of modified gravity with the BI-like Lagrangian density is formulated. A bound on the parameter $\lambda$ with the dimension
(length)$^2$ is obtained. Static solutions corresponding to de Sitter phase without matter are found in Sec.3. In Sec.4, I describe FRW cosmology when the Universe accelerates. The scalar-tensor form of the model is considered and the potential of the scalar field is obtained in Sec.5. It is demonstrated that the de Sitter space is unstable and a solution with zero curvature scalar is stable. The cosmological scenario is described. In Sec.6 the matter stability in the model is investigated and it is shown that there is no matter instability in the proposed model. The summary of results obtained is presented in conclusion.

I use the Minkowski metric of the form $\eta_{\mu\nu}$=diag(-1, 1, 1, 1) and assume that the speed of light $c$ and Planck's constant $\hbar$ to be unity. Greek indices run 0, 1, 2, 3 and Latin indices run the spatial values 1, 2, 3.

\section{The Model of the modified gravitational theory}

Let us consider $F(R)$-gravity with the Lagrangian density
\begin{equation}
{\cal L}=\frac{1}{2\kappa^2}F(R)=\frac{1}{2\kappa^2}\frac{1}{\lambda}
\left(1-\sqrt{1-2\lambda R}\right),
\label{1}
\end{equation}
where $\kappa=\sqrt{8\pi}M_P^{-1}$, $M_P$ is the Planck mass. Some variants of BI-type gravity were considered in \cite{Deser}, \cite{Ketov}, \cite{Wohlfarth}, \cite{Nieto}, \cite{Comelli}, \cite{Comelli1}, \cite{Ferreira}, \cite{Tekin}, \cite{Pani}, \cite{Herdeiro}.
It should be mentioned that the model based on the Lagrangian density (1) is just one of many, because there is no good motivation for any BI-type gravity, unlike the gauge BI action in electrodynamics and string theory.
It is obvious from equation (1) that the scalar curvature has to obey the restriction $R\leq 1/(2\lambda)$. One can say that the smallest size of the universe (during the Big Bang) can not be less than $\sqrt{2\lambda}$. Implying that the constant $\lambda$ with the dimension of (length)$^2$ is small, $\lambda R\ll 1$, we obtain from (1) the Taylor series
\begin{equation}
F(R)=R+\frac{1}{2}\lambda R^2+\frac{1}{2}\lambda^2 R^3 +....
\label{2}
\end{equation}
Thus, at small value of the constant $\lambda$ introduced, one comes to the EH action
$S=1/(2\kappa^2)\int d^4x\sqrt{-g}R$ ($g$=det$g_{\mu\nu}$), as $\lim_{\lambda\rightarrow 0}F(R)=R$. Because GR passes local tests the model under consideration also satisfies observational data at the definite bound on $\lambda$. The laboratory bound from the  E\"{o}t-Wash experiment \cite{Kapner}, \cite{Jetzer} (see also \cite{Berry}, \cite{Zhuk}) gives $F''(0)\leq 2\times 10^{-6}$ cm$^2$. From equation (1), we obtain a restriction on the parameter $\lambda$:
\begin{equation}
\lambda\leq 2\times 10^{-6} cm^2.
\label{3}
\end{equation}
The Taylor series (2) contains all powers in Ricci curvature $R=g^{\alpha\beta}R_{\alpha\beta}$ ($R_{\alpha\beta}$ is the Ricci tensor). At small $\lambda$ series (2) gives the approximate Lagrangian density ${\cal L}=\frac{1}{2\kappa^2}\left(R+\frac{1}{2}\lambda R^2\right)$ which was already considered in \cite{Starobinsky}. Such model results in the self-consistent inflation. $R^2$-term in this Lagrangian prevents from the singular behavior in the past and in the future \cite{Appleby}.

Adding to (1) the Lagrangian
of the matter which is the perfect fluid with the energy-momentum tensor $T^{(m)}_{\alpha\beta}= (p^{(m)}+\rho^{(m)})u_\alpha u_\beta+p^{(m)}g_{\alpha\beta}$ ($p^{(m)}$ is a pressure, $\rho^{(m)}$ is the energy density, and the four-velocity of the fluid obeys $u^\alpha u_\alpha=-1$), one obtains equations of motion (see, for instance, \cite{Faraoni})
\begin{widetext}
\begin{equation}
R_{\mu\nu}F'(R)-\frac{1}{2}g_{\mu\nu}F(R)+g_{\mu\nu}g^{\alpha\beta}\nabla_\alpha\nabla_\beta F'(R) -\nabla_\mu\nabla_\nu F'(R)=\kappa^2T^{(m)}_{\mu\nu},
\label{4}
\end{equation}
\end{widetext}
where $\nabla_\mu$ is a covariant derivative, $F'(R)=dF(R)/dR$. The
conservation of the energy-momentum tensor $\nabla^\mu T^{(m)}_{\mu\nu} = 0$ gives the equation
(for Friedmann$-$Robertson$-$Walker (FRW) metric) as follows:
\begin{equation}
\dot{\rho}^{(m)} + 3H\left(\rho^{(m)} + p^{(m)}\right)=0,
\label{5}
\end{equation}
where $H = \dot{a}(t)/a(t)$ is the Hubble parameter, a(t) is a scale factor and a dot above the variable denotes the differentiation with respect to the time. At the particular case when the equation of state (EoS) parameter $w=p^{(m)}/\rho^{(m)}=-1$, corresponding to the fluid with the property of the dark energy, the energy density $\rho^{(m)}$ becomes constant.

\section{Static Solutions}

Now, we solve equation (4) for Lagrangian density (1) in a particular case when the Ricci scalar is a constant $R=R_0$.
One can add matter with the EoS $w=-1$ introducing a cosmological constant
(a cosmological constant can be included into $F(R)$).
Then all solutions with $R = const$ are given by roots of the algebraic
equation $RF'(R)=2F(R)$ \cite{Barrow}. For the function (1), we obtain
\begin{equation}
\frac{1}{2\lambda}\left(1-\sqrt{1-2\lambda R_0}\right)-\frac{R_0}{4\sqrt{1-2\lambda R_0}}=0.
\label{6}
\end{equation}
Solutions to equation (6) are given by
\begin{equation}
R_0=0,~~~~or~~~~R_0=\frac{4}{9\lambda}.
\label{7}
\end{equation}
We notice that for our model $F'(R)=\left(1-2\lambda R\right)^{-1/2}>0$ and the regime of antigravity $F'(R)<0$ is not realized. In addition, the condition of classical stability $F''(R)\geq 0$ leads to $F''(R)=\lambda(1-2\lambda R)^{-3/2}\geq 0$, i.e. the constant $\lambda$ is positive, $\lambda>0$. For the stability of the Schwarzschild black holes, the weaker condition $F''(0)>0$ would be
sufficient \cite{Appleby}. The positive solution (7) can describe primordial and present dark energy
which is future stable if $F'(R_0)/F''(R_0)>R_0$ \cite{Schmidt}. We find that
\begin{equation}
\frac{F'(R)}{F''(R)}=\frac{1-2\lambda R}{\lambda}.
\label{8}
\end{equation}
Thus, the solution with the flat space-time, $R_0=0$, is stable, $F'(0)/F''(0)>0$, but the solution (7) corresponding to the de Sitter space, $R_0=4/(9\lambda)$ is unstable as $F'(R_0)/F''(R_0)=1/(9\lambda)<R_0$.
We have probably the unification of early-time inflation with late-time acceleration.

Let us consider the spherically symmetric metric with the Schwarzschild form
\begin{equation}
ds^2=-B(r)dt^2+\frac{dr^2}{B(r)}+r^2\left(d\theta^2+\sin^2\theta d\phi^2\right).
\label{9}
\end{equation}
Then for EoS $p^{(m)}=-\rho^{(m)}$ and constant Ricci scalar $R_0$, all $F(R)$ theories admit
Schwarzschild$-$(anti-)de Sitter solutions with the function
\begin{equation}
B(r)=1-\frac{2MG}{r}-\frac{R_0}{12}r^2,
\label{10}
\end{equation}
where $M$ is the mass of the black hole and $G=\kappa^2/(8\pi)$ is Newton's constant. If $R_0>0$, we have the de Sitter space and for $R_0<0$, one has the anti-de Sitter space. According to equation (7) the solutions for the Ricci scalar are positive $R_0\geq 0$ and, therefore, in our model for a constant curvature the de Sitter space is realized. For a free space, without the matter ($p^{(m)}=0$, $\rho^{(m)}=0$), the non-trivial solution (7) $R_0=4/(9\lambda)$ gives the function (10):
\begin{equation}
B(r)=1-\frac{2MG}{r}-\frac{1}{27\lambda}r^2.
\label{11}
\end{equation}
Thus, the classical stability for Schwarzschild black holes leads again to $\lambda>0$. Comparison of solution (11) with the solution of Einstein's equation with cosmological constant $\Lambda$ leads to equality $\Lambda=1/(9\lambda)$. This means that the model suggested mimics the cosmological constant (or the dark energy) for the space without any matter. This is the common property of $F(R)$ models \cite{Barrow}, \cite{Carroll}.

The entropy $S$ in $F(R)$-gravity is given by \cite{Akbar}, \cite{Gong}, \cite{Brustein}
\begin{equation}
S =\frac{F'(R)A}{4G},
\label{12}
\end{equation}
where $A$ is the area of the horizon. From equation (1), one finds
\begin{equation}
S =\frac{A}{4\sqrt{1-2\lambda R}G}.
\label{13}
\end{equation}
It follows from (13) that instead of Newton constant $G$, one can introduce the effective gravitational coupling $G_{eff}=\sqrt{1-2\lambda R}G$. For the nontrivial solution (7) ($R_0=4/(9\lambda)$), we arrive at the effective gravitational constant $G_{eff}=G/3$.

\section{FRW cosmology}

In homogeneous, isotropic and spatially flat FRW cosmology
the space-time metric is given by
\begin{equation}
ds^2 = -dt^2 + a^2(t)\left(dx^2+dy^2+dz^2\right).
\label{14}
\end{equation}
The scalar curvature R is expressed via the the Hubble parameter as follows: $R = 12H^2 + 6\dot{H}$.
In this case equation (4) reduces to a system of two equations
\begin{widetext}
\begin{equation}
\frac{F(R)}{2}-3\left(H^2+\dot{H}\right)F'(R)+18\left(4H^2\dot{H}+ H\ddot{H}\right)F''(R)=\kappa^2\rho^{(m)},
\label{15}
\end{equation}
\begin{equation}
\frac{F(R)}{2}-\left(3H^2+\dot{H}\right)F'(R)+6\left(8H^2\dot{H}+4\dot{H}^2+6H\ddot{H}+
\dddot{H}\right)F''(R)+36\left(4H\dot{H}+\ddot{H}\right)F'''(R)=-\kappa^2p^{(m)}.
\label{16}
\end{equation}
\end{widetext}
For solutions $\dot{H}_0=0$, these two equations (19) are consistent with EoS $\rho^{(m)}=-p^{(m)}$. We consider here the model without any matter. Then $H_0=\sqrt{R_0/12}$ and from equation (7), one obtains $H_0=1/(3\sqrt{3\lambda})$ for a de Sitter phase. As a result a scale factor becomes
\begin{equation}
a(t)=a_0\exp\left(\frac{t}{3\sqrt{3\lambda}}\right)
\label{17}
\end{equation}
and describes the inflation phase. It should be mentioned that Eq.(17) describes the eternal inflation (i.e. has no end) and thus it is not viable, unless some means to its end are supplied. To have the detailed description of inflation, one has to obtain the exact solution of Eqs.(15),(16) for the general case $R\neq const$.

\section{The Scalar-Tensor Form of the Theory}

\begin{figure}[h]
\includegraphics[height=2.0in,width=3.0in]{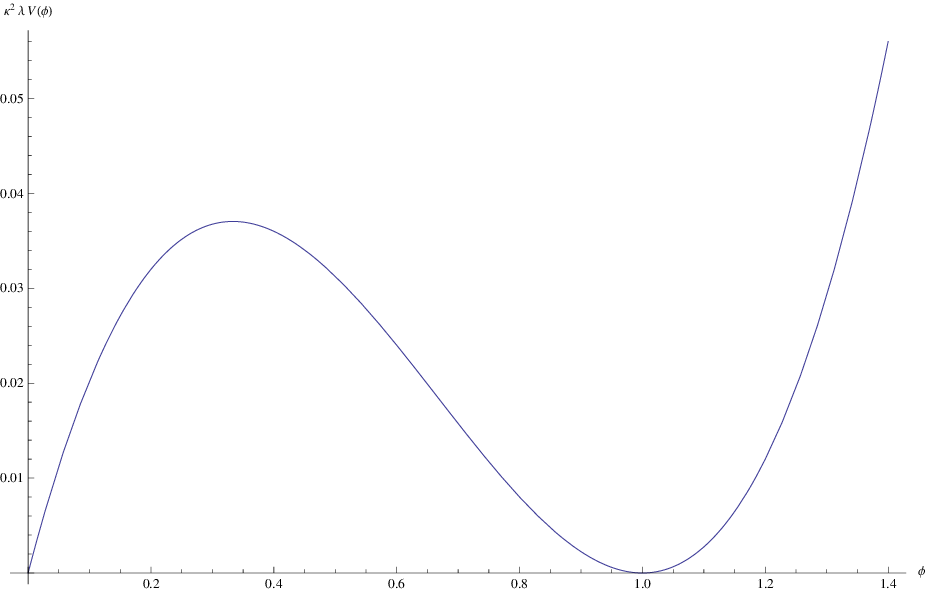}
\caption{\label{fig.1}$V(\phi)\lambda\kappa^2$ versus $\phi$. There is maximum at $\phi=1/3$ and minimum at $\phi=1$.}
\end{figure}

There is a link between modified $F(R)$-gravity and scalar-tensor theories of gravity. Equation (1) presents modified gravity in the Jordan frame with metric tensor variables $g_{\mu\nu}$. Let us consider the Einstein frame with conformally transformed metric \cite{Sokolowski}
\begin{equation}
\widetilde{g}_{\mu\nu} =F'(R)g_{\mu\nu}=\frac{g_{\mu\nu}}{\sqrt{1-2\lambda R}}.
\label{18}
\end{equation}
Then, in new variables (18), equation (1) becomes
\begin{equation}
{\cal L}=\frac{1}{2\kappa^2}\widetilde{R}-\frac{1}{2}\widetilde{g}^{\mu\nu}
\nabla_\mu\varphi\nabla_\nu\varphi-V(\varphi),
\label{19}
\end{equation}
where $\widetilde{R}$ is defined in new metric (18). We have introduced the scalar field $\varphi$ and the potential $V(\varphi)$ as follows:
\begin{equation}
\varphi=-\frac{\sqrt{3}}{\sqrt{2}\kappa}\ln F'(R)=\frac{\sqrt{3}}{\sqrt{2}\kappa}\ln \sqrt{1-2\lambda R},
\label{20}
\end{equation}
\begin{equation}
V(\varphi) =\frac{RF'(R)-F(R)}{2\kappa^2F'^2(R)}|_{R=R(\varphi)}=\frac{\phi\left(1-\phi\right)^2}{4\lambda\kappa^2},
\label{21}
\end{equation}
where $\phi=\exp\left(\sqrt{2}\varphi \kappa/\sqrt{3}\right)=1/F'(R)$. The plot of the function $V(\phi)$ (21) is presented in Fig.\ref{fig.1}. One can also plotting the graph $V(\varphi)$ since $\varphi$ is the canonical scalar field.
The potential function (21) possesses minimum at $\phi=1$ ($V''(\phi=1)>0$) and maximum at $\phi=1/3$ ($V''(\phi=1/3)<0$). We find that for static solutions (7) the value $R_0=0$ corresponds to the minimum and $R_0=4/(9\lambda)$ corresponds to the maximum of the potential. Thus, again we make a conclusion that  the zero scalar curvature is the stable state and the state with Ricci scalar $R_0=4/(9\lambda)$ is unstable. From potential (21), we obtain the mass of a scalar state
\begin{equation}
m_\varphi^2=\frac{d^2V}{d\varphi^2} =\frac{\phi\left(9\phi^2-8\phi+1\right)}{6\lambda}.
\label{22}
\end{equation}
It follows from (22) that $m_\varphi^2=1/(3\lambda)>0$ for $R_0=0$ ($\phi=1$) and $m_\varphi^2=-1/(27\lambda)<0$ for $R_0=4/(9\lambda)$ ($\phi=1/3$). The solution with constant curvature $R_0=4/(9\lambda)$ leads to unstable de Sitter space (9), (11), but the solution with zero scalar curvature $R_0=0$ gives the condition $m_\varphi^2>0$ for a stability of the Schwarzschild solution (10) ($R_0=0$). The criterion of the stability of the de Sitter solution in F(R) gravity was first obtained in \cite{Schmidt}.  As the constant $\lambda$ is small the squared mass $m_\varphi^2$ is big and corrections to the Newton law are negligible.

One can describe the  cosmological scenario as follows. The Universe starts with the large finite positive curvature $R=1/(2\lambda)$ (because the parameter $\lambda$ is small) corresponding to the field $\phi=0$ and it is apparently analogous to $\varphi=-\infty$. It should be mentioned that classical F(R)-gravity can not describe precisely this trans-Planckian region.  For the large positive curvature quantum gravity corrections should be taken into account. Then the Universe inflates and has the positive curvature $R_0=4/(9\lambda)$ being in de Sitter's phase. This state is unstable and the Universe rapidly expands with the Hubble parameter $H_0=\sqrt{R_0/12}$ (for the constant $R_0$) which defines the expansion rate $\dot{a}(t)/a(t)=H_0$ of the de Sitter space. Then the scalar curvature decreases (rolls down) and becomes small so that the rate of the expansion $H$ is small. The final state is stable and corresponds to the vanishing Ricci curvature $R_0=0$. Thus, the Universe approaches a stable Minkowski space.

It should be mentioned that F(R) model under consideration is not suitable for the inflationary scenario of the early Universe since the de Sitter regime (17) in it is too unstable,
i.e. $|m^2|$ in the physical, Jordan frame, is not small compared to $H_0^2=1/(27\lambda)$.

\section{Matter Stability}

Now we consider the equation of motion for a curvature scalar. Taking the trace of left and right sides of equation (4), one obtains
\begin{equation}
3g^{\alpha\beta}\nabla_\alpha\nabla_\beta F'(R)+ F'(R)R-2F(R)=\kappa^2T^{(m)},
\label{23}
\end{equation}
where $T^{(m)}=T^{(m)}_{\mu\nu}g^{\mu\nu} $.
To investigate the matter stability, we follow \cite{Dolgov}, and apply equation (23) for weak gravity objects. The flat Minkowski metric can be used for weak gravitation so that $g^{\alpha\beta}\nabla_\alpha\nabla_\beta\simeq \partial_k^2-\partial_t^2$.
For spatially constant distribution ($R$ is uniform) equation (23) becomes
\begin{equation}
-3F^{(2)}(R)\ddot{R}-3F^{(3)}(R)\dot{R}^2+ F^{(1)}(R)R-2F(R)=\kappa^2T^{(m)},
\label{24}
\end{equation}
where $F^{(n)}(R)=d^nF(R)/dR^n$.
We consider a perturbative solution $R=R_0+R_1$ ($R_1$ is the perturbed part, $|R_1|\ll|R_0|$), where in the lowest order the curvature, according to GR, is $R_0 =- \kappa^2T^{(m)}$ inside the matter and $R_0 = 0$ outside the matter. From equation (24), one obtains (see \cite{Odintsov},
\footnote{I have corrected typos in \cite{Odintsov}.})
\begin{widetext}
\begin{equation}
\ddot{R}_0+\ddot{R}_1+\frac{F^{(3)}(R_0)}{F^{(2)}(R_0)}\left(\dot{R}_0^2+2\dot{R}_0\dot{R}_1\right)+ \frac{2F(R_0)-R_0\left[1+F^{(1)}(R_0)\right]}{3F^{(2)}(R_0)}=U(R_0)R_1,
\label{25}
\end{equation}
where
\begin{equation}
U(R_0)=\frac{F^{(3)2}-F^{(2)}F^{(4)}}{F^{(2)2}}\dot{R_0}^2+ \frac{\left(R_0F^{(2)}-F^{(1)}\right)F^{(2)}+\left(2F-R_0F^{(1)}-R_0\right)
F^{(3)}}{3F^{(2)2}}.
\label{26}
\end{equation}
\end{widetext}
The system is unstable if $U(R_0)>0$ because $R_1$ exponentially increases in the time. We find that for our model (Eq. (1)) the function $U(R_0)$ is given by
\begin{equation}
U(R_0)=-\frac{6\lambda^2\dot{R_0}^2}{\left(1-2\lambda R_0\right)^2}
+ \frac{12\lambda R_0-7+3\left(2-\lambda R_0\right)\sqrt{1-2\lambda R_0}}{3\lambda}.
\label{27}
\end{equation}
One can verify that for static solutions (7) $U(R_0)<0$ that indicates on stability of the system. Even for the least value of the scalar curvature $R_0=1/(2\lambda)$, we have $U(R_0)<0$. As a result, there is no matter instability and the model passes the matter stability test.

\section{Conclusion}

A modified theory of gravity with the Born$-$Infeld-like action is
suggested and analyzed. At limiting case $\lambda\rightarrow 0$ the action introduced is converted into
EH action. The Jordan frame as well as the Einstein frame were considered and the potential of the scalar field in scalar-tensor form of the theory was obtained and is presented in Fig.1. We have found the static Schwarzschild$-$de Sitter solutions of the model with the de Sitter space to be unstable and the solution with zero Ricci scalar to be stable. The constant $\lambda$ introduced can be connected with the fundamental length $l=\sqrt{2\lambda}$ so that the smallest size of the Universe during the Big Bang is $l$. From the local tests bound on the constant is $\lambda\leq 2\times 10^{-6}$ cm$^2$, and we obtain the restriction on the fundamental length $l\leq 2\times 10^{-3}$ cm. It was demonstrated that no matter instability in the model suggested. There are open questions in the model considered which I leave for further investigations: is the very low bound for $l$ really compatible with all observations, including particle physics?; to obtain exact solutions describing inhomogeneities in a FRW universe, etc.


\end{document}